\documentclass[aps,prc,twocolumn,superscriptaddress,nofootinbib,showpacs,showkeys,preprintnumbers]{revtex4-1}
\usepackage{graphicx}  
\usepackage{bm}        
\usepackage{amssymb}   
\usepackage{amsmath}   
\usepackage{hyperref}
\usepackage{multirow}
\usepackage{lineno}
\usepackage{comment}
\usepackage{subcaption}
\usepackage{subfiles}

\newcommand{\feqn}{f_{\mathrm{eq,n}}}
\newcommand{\feqnb}{\bar f_{\mathrm{eq,n}}}
\newcommand{\cJ}{{\cal J}}
\newcommand{\cN}{{\cal N}}
\newcommand{\cM}{{\cal M}}
\newcommand{\Peq}{{\cal P}_{\mathrm{eq}}}
\newcommand{\cG}{{\cal G}}
\newcommand{\cF}{{\cal F}}

\begin{document}

\title{Elliptic flow of deuterons in ultrarelativistic heavy-ion collisions}

\author{Radka Voz\'abov\'a} \affiliation{Faculty of Nuclear Sciences and Physical Engineering, Czech Technical University in Prague,\\  B\v rehov\'a 7, 11519 Prague 1, Czech Republic}
\author{Boris Tom\'a\v{s}ik} \affiliation{Faculty of Nuclear Sciences and Physical Engineering, Czech Technical University in Prague,\\  B\v rehov\'a 7, 11519 Prague 1, Czech Republic}
\affiliation{Univerzita Mateja Bela, Tajovsk\'eho 40, 974~01 Bansk\'a Bystrica, Slovakia}

\begin{abstract}
We calculate the elliptic flow of deuterons in Pb+Pb collisions at 2.76~TeV per colliding nucleon-nucleon pair and show that it can be used to discriminate between direct statistical production and coalescence.
The emission from the fireball is parametrized and tuned to reproduce transverse momentum spectra and the elliptic flow of protons and pions. 
Coalescence leads to higher deuteron elliptic flow than statistical production and agrees better with experimental data.
We attribute this observation to the varying size of the nucleon-producing region for the emission in different azimuthal angles. 
\end{abstract}
\maketitle


\section{Introduction and motivation}
\label{s:intro}

Ultrarelativistic collisions of heavy atomic nuclei are examined with the aim to probe matter at highest temperatures ever created in laboratory. However, a few very interesting results have been achieved in parallel to the main line of research. One of them is the production of light nuclear clusters. Even very exotic ones have been identified among the produced particles, like antimatter $^4\overline{\mbox{He}}$  nucleus \cite{STAR:2011eej} or the hypertriton as well as anti-hypertriton \cite{STAR:2019wjm}. Nevertheless, while these measurements were crucial for the understanding of the properties of the measured clusters, the jury is still inconclusive about the mechanism of their production. 

The inconclusivity occurs already at the level of the simplest nuclear clusters---deuterons and antideuterons. On one side, the yields of deuterons, antideuterons, and heavier clusters alongside with other types of hadrons including pions, kaons, and protons can all be  reproduced by the Statistical Model \cite{Andronic:2017pug,Andronic:2016nof} reasonably well. It seems as if the Statistical Model really provided a universal description for the bulk properties of the matter reflected in hadron species and their clusters. On the first sight, however, this is extremely puzzling, since the cluster binding energy is two orders of magnitude  below the characteristic temperature, the former being a few MeV and the latter about 150~MeV. Quite naturally, production of deuterons from a system at such a high temperature has been coined by the metaphor 'snowballs in hell` \cite{Andronic:2017pug}.

The results \cite{Zhu:2017zlb, Andronic:2017pug} show that both the statistical model and the coalescence mechanism predict similar deuteron yields. However, when using a blast-wave model as a representation of thermal production of the proton $p_t$ spectra and $v_2$ and simply replacing the proton mass by those of light nuclei, the data on the elliptic flows of these light nuclei \cite{STAR:2016ydv} are not well reproduced.

Coalescence is a natural mechanism for the deuteron and cluster production \cite{Butler:1963pp,Gutbrod:1976zzr,Sato:1981ez,Gyulassy:1982pe,Csernai:1986qf,Mrowczynski:1987oid,Lyuboshits:1988yc,Scheibl:1998tk,Chen:2012us,Chen:2013oba,Sun:2015jta,Dong:2018cye,Sun:2018jhg,Sombun:2018yqh,Hillmann:2021zgj}. 
Due to their low binding energy it is assumed that the clusters are formed by coupling nucleons only after the latter have escaped the strongly interacting fireball. 

The similarity of results from statistical  production and coalescence is intriguing and it provokes a question if such an agreement of the different models is a robust feature based on underlying physics processes or mere coincidence resulting from fine tuning of a number of small contributing effects \cite{Mrowczynski:2016xqm}. Even though a bound cluster cannot really be produced in a hot environment, an idea was put forward about an existence of correlated nucleon n-tuples already at high temperatures, that could later transform themselves into observed clusters \cite{Andronic:2017pug}.

Description of deuteron production in dynamical models goes also long time back \cite{Danielewicz:1991dh}. Current data are addressed in dynamical studies with transport models which focus on the microscopic production and survival of deuterons \emph{within} the hot hadronic phase. Studies with SMASH \cite{Oliinychenko:2018ugs,Staudenmaier:2021lrg} included destroying and regenerating scattering of deuterons, but ignored their very large formation time \cite{Mrowczynski:2020ugu}. A newly constructed Parton-Hadron Quantum-Molecular Dynamics model (PHQMD) \cite{Aichelin:2019tnk} includes direct interaction between baryons beyond the mean field and a cluster finding algorithm that identifies the clusters during the evolution of the hot hadronic phase. Experimental results on rapidity distributions and transverse momentum spectra were successfully reproduced for energies of few hundreds MeV up to 200~GeV per colliding NN pair \cite{Glassel:2021rod,Kireyeu:2022qmv,Coci:2023daq,Kireyeu:2023bye}. It was shown that coalescence, which is applied only after the breakup of the hadronic phase, yields similar results for the deuterons \cite{Kireyeu:2022qmv}.
Kinetic approach to production of lighter nuclei has also been applied for nucleon reactions at intermediate energies \cite{Wang:2023gta}.

Due to the finite size of the deuteron, coalescence depends on the spatio-temporal extension of the source, which is measured by femtoscopy \cite{Mrowczynski:1987oid,Mrowczynski:1989jd,Mrowczynski:1989jd,Mrowczynski:1992gc,Mrowczynski:1993cx}. 
To more explicitly include this feature, quantum-mechanical formalism of \cite{Scheibl:1998tk} has recently been generalised \cite{Blum:2019suo}.
Another extension, motivated also by the need to explain antideuteron production in collisions of cosmic ray particles, which are usually small systems, has also been worked out \cite{Kachelriess:2019taq,Kachelriess:2020amp,Kachelriess:2023jis} and implemented for computation \cite{Kachelriess:2022khq}.

The constraint of coupling coalescence to the source sizes has been implemented in several analyses that aim at distinguishing the true production mechanism. Early simple estimates of the coalescence parameter \cite{Bellini:2018epz} were later improved with more advanced formalism \cite{Blum:2019suo,Bellini:2020cbj} to the extent that the model deuteron wave function could be optimised to fit the experimental data best \cite{Mahlein:2023fmx}.

Other possibilities to discriminate between the two production models include examining the production of very exotic spatially extended and high spin isotope $^4$Li \cite{Bazak:2020wjn}, and/or performing femtoscopic studies involving correlations of deuterons to protons, another hadrons \cite{Mrowczynski:2019yrr}, or among themselves \cite{Mrowczynski:2021bzy}.

Recently, an enhancement of deuteron production within jet cones \cite{ALICE:2020hjy,ALICE:2022ugx} was also successfully interpreted within the coalescence picture \cite{Mrowczynski:2023hbn}.

In \cite{Feckova:2016kjx}, the cumulants of the deuteron number distribution were argued to be enhanced for coalescence with respect to the prediction from the Statistical Model. Measurements from ALICE  \cite{ALICE:2022xiu} and STAR \cite{STAR:2023egt} collaborations indicate a disagreement with the predictions from coalascence. 

In this paper, we examine the idea to use $v_2$ measurements of deuterons for the distinction between coalescence and statistical production. 
The elliptic flow of deuterons from coalescence has been simulated before \cite{Oh:2009gx,Yin:2017qhg,Zhu:2017zlb,Zhao:2018lyf,Zhu:2015voa}
but the comparison of the models has not been done. 
By doing it, we further dwell on the idea  that deuterons from coalescence are a femtoscopic probe., i.e., their production is determined by the size of the homogeneity region from which their nucleons were produced. This is mostly the case if the homogeneity region is comparable to the typical size of the deuteron wave function, which is about 2~fm. Based on this, we suppose that the elliptic flow of deuterons may be a sensitive test of coalescence, because it measures the variations of deuteron yield in different azimuthal directions. Deuterons produced in different azimuthal directions are built from nucleons that originate from corresponding different homogeneity regions, which in non-central collisions may differ by size. This may lead to the sensitivity of coalescence to the direction of deuteron production. 

In a way, we propose to look at more detailed structure of the source than was done so far in studies of deuteron production. This will require to devote more attention to the modelling of the source, which we will constrain from hadronic observables. 

For the actual calculation we use a model with parametrized freeze-out hypersurface, included production of hadrons from decays of resonances, and implemented corrections to the momentum distribution function due to viscosity. The parametrization is based on an extension and upgrade of the blast-wave model. Our results indicate---in accordance with the reasoning above---that thermal production and coalescence indeed lead to different elliptic flows, provided that the fireball anisotropy is large enough. 

We explain this approach in more detail in Section \ref{s:approach}. The model used here is introduced in Section~\ref{s:model}, together with its calibration. Our results are summarised in Section~\ref{s:res} and we conclude in Section~\ref{s:conc}.
Technical description of the freeze-out implementation is placed in the Appendix.


\section{The approach}
\label{s:approach}

In this paper we are interested in the  mechanism of deuteron production, which happens at the kinetic freeze-out. We will not try to link that production to previous evolution of the bulk matter, its properties and/or initial conditions. The reason is that we chose to look at differential $v_2$, which is statistics-hungry in simulations. Thus, getting it from a dynamical simulation for various setups would easily become practically intractable. Also, we want have the freedom to explore the production process at carefully chosen  freeze-out while not making the approach more complicated than necessary. Therefore, we decided to use a \emph{parametrization} of the kinetic freeze-out state of hadrons which we can tune with the help of a few parameters.

A representation of such a parametrization is the well-known blast-wave model. Here, we extend this model further in several aspects, guided by experimental data. 

In the hadronic fireball, just before its kinetic freeze-out, nucleons and pions interact strongly with each others. In fact, pions are the mediators of nucleon-nucleon interaction. This leads us to assume that they decouple from the fireball at the same time and must be described by the same emission function, with a different mass inserted. This sets the constraint for our modelling of the kinetic freeze-out: we require that the model reproduces azimuthally integrated $p_t$-spectra of protons and charged pions, as well as their $v_2$ as function of $p_t$. During our work, it has turned out that the requirement to fit all these data of the two identified species fixes the parametrization rather uniquely. 

Then, the subsequent calculation of the $p_t$-distribution and the differential $v_2$ of the deuterons can be considered as a prediction, since there is basically no freedom to re-adjust the model, which has been set  on protons and pions.  

We can then directly generate the thermally produced deuterons, we just need to set the mass to that of the deuteron and use proper spin degenaracy factor. 

Coalescence is simulated in two steps. The mechanism can be understood  from the formula for the single-particle spectrum of deuterons \cite{Scheibl:1998tk}
\begin{multline}
    E_d \frac{dN_d}{d^3P_d} = \frac{3}{8(2\pi)^3} \int_{\Sigma_f} P_d \cdot d\Sigma_f(R_d)
    \\ 
    f_p\left (R_d, \frac{P_d}{2}\right )
    f_n\left (R_d, \frac{P_d}{2}\right )
    {\cal C}_d(R_d,P_d)
\label{e:Dspect}
\end{multline}
where we use $a\cdot b = a_\mu b^\mu$ for the scalar product and the factor ${\cal C}_d(R_d,P_d)$ represents 
\begin{multline}
{\cal C}_d(R_d,P_d) = 
\\
\int \frac{d^3q d^3r}{(2\pi)^3}  
\frac{f_p(R_+,P_+)f_n(R_-,P_-)}{f_p(R_d,P_d/2)f_n(R_d,P_d/2)} 
{\cal D}(\vec r,\vec q)\,  .
\label{e:QMcorr}
\end{multline}
The integration in Eq.~(\ref{e:Dspect}) runs over the kinetic freeze-out hypersurface, $d\Sigma_f(R_d)$ is surface element along the hypersurface, and $R_d$ marks a position on that hypersurface, ascribed to the deuteron. Distributions of protons and neutrons in phase-space are denoted $f_p$ and $f_n$. The actual mechanism of coalescence is represented by the factor ${\cal C}_d(R_d,P_d)$.  The integration in Eq.~(\ref{e:QMcorr}) runs over relative position $\vec r$ and relative momentum $\vec q$ of the nucleon pair, in the rest frame of the deuteron, and $R_{+}$, $R_-$, $P_+$, $P_-$ are their positions and momenta in that frame. Finally, ${\cal D}(\vec r,\vec q)$ stands for the deuteron Wigner density
\begin{equation}
{\cal D}(\vec r,\vec q) = \int d^3\xi e^{-i\vec q\cdot \vec \xi}
\varphi_d\left (\vec r + \frac{\vec \xi}{2} \right )
\varphi_d^*\left (\vec r - \frac{\vec \xi}{2} \right )\, ,
\end{equation}
where $\varphi_d$ is the deuteron wave function. 

Hence, deuteron yield is given by the convolution of the deuteron Wigner function with the distributions (Wigner functions) of the protons and neutrons. Those distributions will be provided by our model, tuned on protons and pions. The actual Monte Carlo application will be explained in the next Section. 

Equation \ref{e:Dspect} allows to better illustrate why deuteron is a femtoscopic probe. For nuclear collisions, we will assume that the nucleon distributions vary with momentum much slower than the deuteron Wigner density, so that the integration over momentum difference in Eq.~(\ref{e:QMcorr}) is limited by ${\cal D}(\vec r,\vec q)$. 

If the homogeneity region for nucleons with given momentum is much larger than the typical spatial extension of the deuteron wave function, then the spatial integration in Eq.~(\ref{e:QMcorr}) covers the whole wave function. Due to its normalization, the whole integral tends to 1. In this case, deuteron production is equivalent to the production of one proton and one neutron, multiplied by the appropriate degeneracy factor. 

On the other hand, if the emitting region is much smaller than the size of the deuteron wave function, then the integration in Eq.~(\ref{e:QMcorr}) is limited by $f_p\cdot f_n$ and gives a factor smaller than 1, representing just how big part of the deuteron wave function is included. The interesting regime is where the size of the emitting region is just about the size of the deuteron, because this is when the former behavior changes into the latter. 

Deuteron production thus becomes sensitive to the size of the emitting source and we want to use this dependence in our study.


\section{The model}
\label{s:model}

In Statistical (thermal) Model, particle yields are given by the volume, temperature, and chemical potentials for the conserved charges. 
Together with spatial distributions of the emission points and the profile of the collective expansion velocity these are all ingredients that determine the so-called Blast-Wave (BW) model \cite{Schnedermann:1993ws,Csorgo:1995bi,Tomasik:2004bn}. It will be extended here, however.

\subsection{Freeze-out hypersurface}

We parametrize the momentum of a particle with the help of transverse momentum $p_t$, transverse mass $m_t$, rapidity $Y$ and the azimuthal angle $\phi$ as
\begin{equation}\label{eq:1}
    p^{\mu} = \left(m_t \cosh Y, p_t \cos \phi, p_t \sin \phi, m_t \sinh Y \right).
\end{equation}
In this paper, we shall denote the rapidity with capital letter in order to distinguish it from the spatial coordinate. 

In the position space, we shall use the polar coordinates,  $r$ and the azimuthal angle $\Theta$, in the transverse plane. Hyperbolic coordinates map the remaining two directions:  the space-time rapidity 
\begin{equation}\label{eq:2}
    \eta_s =  \dfrac{1}{2} \ln \left( \dfrac{t + z}{t - z } \right)\,  ,
\end{equation}
and the longitudinal proper time $\tau = \sqrt{t^2 - z^2}$. Then, one can write for the coordinates
\begin{equation}\label{eq:3}
    x^{\mu} = \left( \tau \cosh \eta_s, r \cos \Theta, r \sin \Theta, \tau \sinh \eta_s \right).
\end{equation}

When trying to simultaneously reproduce the identified $p_t$-spectra and $v_2(p_t)$ of protons and pions, we found out that the usual formulation of the blast-wave model with the freeze-out along constant longitudinal proper time hyperbola is insufficient. Thus we replace it with the quadratic dependence of the freeze-out time on the radial coordinate 
\begin{equation}
    \tau_f(r) = s_0 + s_2 r^2\, ,
\label{e:fordep}
\end{equation}
where $s_0$ and $s_2$ are parameters of the model.


The three-dimensional element of the freeze-out hypersurface is given by
\begin{equation}\label{eq:4}
d^3 \Sigma_\mu = \varepsilon _{\mu \nu\kappa\rho} \dfrac{dx^\nu}{dr}\dfrac{dx^\kappa}{d\Theta}\dfrac{dx^\rho}{d\eta_s}dr\, d\Theta\, d\eta_s,
\end{equation}
where $\varepsilon _{\mu \alpha \beta \gamma} $ is the Levi-Civita tensor.

Then,  from Eq.(\ref{eq:3}) one gets
\begin{multline}\label{eq:5}
  d^3\Sigma^\mu = \left( \cosh \eta_s,  \dfrac{\partial \tau_f (r)}{\partial r} \cos \Theta,  \dfrac{\partial \tau_f (r)}{\partial r} \sin \Theta, \sinh \eta_s \right) \\
  \times r \tau_f(r) d \eta_s dr d \Theta\, ,
\end{multline}
and Eq.~(\ref{e:fordep}) leads to 
\begin{multline}
      d^3\Sigma^\mu = \left( \cosh \eta_s,  2s_2 r \cos \Theta,  
      2s_2 r \sin \Theta, \sinh \eta_s \right) \\
  \times r \tau_f(r) d \eta_s dr d \Theta\, ,
\end{multline}
Note that Eqs.~(\ref{eq:3}) and (\ref{eq:5}) just show how the positions in space and the freeze-out hypersurface are parametrised, respectively, with the help of coordinates. They do not yet desribe the actual shape of the hot fireball.  

We further extend the usual BW model to incorporate azimuthal anisotropies both in flow and spatial distribution. The radial size shows second-order oscillation around the mean $R_0$ with an amplitude $a_2R_0$
\begin{equation}\label{eq:7}
R(\Theta) = R_0 \left( 1 - a_2 \cos (2 \Theta) \right)  \,  .
\end{equation}
Technically, $R(\Theta)$ sets the maximum value that the coordinate $r$ can attain. The azimuthal spatial anisotropy of the fireball is set by the $\Theta$-dependence of this maximum. 
The expansion velocity is parametrised as 
\begin{multline}
    u^\mu = (\cosh\eta_s \cosh\rho(r),\, \sinh\rho(r)\cos\Theta_b,\, \\
     \sinh\rho(r)\sin\Theta_b,
    \sinh\eta_s\cosh\rho(r) )\,  ,
\end{multline}
where the transverse rapidity grows linearly with (dimensionless) radial coordinate 
\begin{equation}\label{eq:8}
\bar{r} = r / R(\Theta)  
\end{equation}
and exhibits an azimuthal oscillation parametrized by $\rho_2$
\begin{equation}\label{eq:6}
\rho (\bar{r}, \Theta_b) = \bar{r} \rho_0 \left( 1 + 2 \rho_2 \cos (2 \Theta_b )  \right)\,  ,
\end{equation}
with $\rho_0$ being the average transverse rapidity gradient.
The velocity is directed along the angle 
$\Theta_b = \Theta_b (r, \Theta)$, 
which points into the direction perpendicular to the elliptical surface of the freeze-out
\begin{equation}\label{eq:10}
\tan \left( \Theta_b - \frac{\pi}{2} \right) = \dfrac{dy}{dx} = \dfrac{\dfrac{dy}{d \Theta}}{\dfrac{dx}{d \Theta}},  
\end{equation}
where the derivatives are taken along an $\bar{r} = \mathrm{const}$ curve.

\subsection{Momentum distribution}

Along the freeze-out hypersurface, hadrons are produced according to the Cooper-Frye formula \cite{Cooper:1974mv}
\begin{equation}
E\frac{d^3N}{dp^3} = \int_\Sigma d^3\Sigma\cdot p f(x,p)\,  ,
\label{e:CF}
\end{equation}
where $f(x,p)$ is the distribution of the emission points in position and momentum space. Thermal production of deuterons is also calculated according to this formula. 

In the simplest case, within the local rest frame of a moving fluid, $f(x,p)$ is assumed to be Bose-Einstein or Fermi-Dirac distribution (or their Boltzmann approximation) \cite{Schnedermann:1993ws,Csorgo:1995bi}. 

Nevertheless, if the freeze-out happens from a viscous medium, that distribution must be modified \cite{Teaney:2003kp}. Here, we adopt the parametrization of \cite{McNelis:2021acu} which is based on the approach of \cite{Pratt:2010jt}. It consists of modifying the momenta, temperature and the chemical potential term in the original statistical distribution. The distribution for particles with  mass $m_n$, baryon number $b_n$, and spin degeneracy $g_n$ at temperature $T$ then becomes
\begin{equation}
\label{eq:5.9}
    f_{eq,n}^{PTM} = \dfrac{Z_n g_n}{\exp \left[ \dfrac{\sqrt{\boldsymbol{p}^{\prime 2} + m_n^{2}}}{T + \beta_{\Pi}^{-1} \Pi \cF} - b_n \left( \alpha_B + \dfrac{\Pi \cG}{\beta_{\Pi}}\right) \right] + \Theta_n},
\end{equation}
where $\alpha_B$ is the ratio of baryon chemical potential to temperature $\mu_B/T$, and $\Theta_n$ is 1 (--1) for fermions (bosons). 
Other terms and symbols are explained in Appendix \ref{a:PTM}.

For the actual generation of hadrons we have adopted the HadronSampler from the SMASH package \cite{SMASH:2016zqf,Schafer:2021csj}. The original viscous corrections implemented there have been replaced by Eq.~(\ref{eq:5.9}), because in the original formulation the corrections may outgrow the main term. 

Resonance production is included in our model and SMASH was  used to simulate their decays. However,  subsequent rescatterings of hadrons were forbidden. 

Chemical potentials for the hadrons at freeze-out have been assumed according to the partial chemical equilibrium scenario \cite{Bebie:1991ij}. The chemical freeze-out is assumed to happen at the temperature 156~MeV \cite{Melo:2019mpn}, but the hadronic system stays together longer and cools down until it reaches the thermal freeze-out. 
In partial chemical equilibrium, this leads to the appearance of proper chemical potentials for all stable hadrons. Chemical potentials for  resonances are determined from the condition that their production and decay  keeps them in equilibrium with ground state hadrons. The chemical potentials then depend on temperature. In general, at the chemical freeze-out they coincide with the values determined in chemical equilibrium and then they grow as the temperature decreases. We took their values from \cite{Melo:2019mpn}.


\subsection{Coalescence}
\label{ss:coal}

Nucleons generated by the above procedure may subsequently couple to form a deuteron. In the simulation, this is implemented along the following steps \cite{Sombun:2018yqh}: 
\begin{enumerate}
    \item We consider protons and neutrons at the place of their last interaction with the fireball or where their parent resonance decayed (i.e., their freeze-out).
    \item For each p-n pair, the momentum and position of the proton and neutron is boosted to the center-of-mass frame  of the pair.
    \item The particle that has decoupled earlier is propagated  to the time when the other particle was created.
    \item We calculate the relative momentum $\Delta p = |\boldsymbol{p_1} - \boldsymbol{p_2}|$ and the relative distance $\Delta r = |\boldsymbol{r_1} - \boldsymbol{r_2}|$ at equal times  of the p-n pair in their center-of-mass frame.
    \item To create a deuteron, we require that $\Delta p < \Delta p_{max}$ and $\Delta r < \Delta r _{max}$.
    \item Next, the statistical spin and isospin factor ${3}/{8}$ due to the spin and isospin projection to the deuteron state (probability $1/2 \cdot 3/4 = 3/8$) is added. 
    \item Finally, the chosen p-n pair is marked as a deuteron and its proton and neutron are removed from the list of created hadrons.
\end{enumerate}


\subsection{Model calibration}

Let us recall our goal, which is to test if the two potential production mechanisms for deuterons can be discriminated by their prediction for $v_2(p_t)$.  The $p_t$ spectra of deuterons must be reproduced, but we do not aim at any resolving power here, for the production mechanism. 

In this study we plan to calculate $v_2$ for one semi-central, one mid-central, and one peripheral centrality class. Data on $v_2(p_t)$ of deuterons were published in \cite{ALICE:2017nuf} for centrality classes 0-5\%, 5-10\%, 10-20\%, 20-30\%, 30-40\%, 40-50\%. We chose 0-5\% and 30-40\% for the comparison with data and 50-60\% for a prediction, where the effects may be strongest due to higher degree of anisotropy, even though no experimental data are available there. 

Unfortunately, the $p_t$ spectra of deuterons have been published in different classes of centrality: 0-10\%, 10-20\%, 20-40\% \cite{ALICE:2017nuf}. We will take this into account when simulating them. 

For each centrality class, parameters of our model are tuned on $p_t$ spectra and $v_2(p_t)$ for protons and pions. The spectra were published in \cite{ALICE:2013mez} and the $v_2$ in \cite{ALICE:2014wao}, 
and they were divided into centrality classes that overlap with $v_2(p_t)$ for deuterons in \cite{ALICE:2017nuf}.

Proceeding with the calibration of the model we note that the azimuthally integrated $p_t$ spectra do not depend on the anisotropy parameters $a_2$ and $\rho_2$. This allows to set the temperature $T$ and transverse flow,  parametrized with $\rho_0$, from the comparison to $p_t$ spectra of identified hadrons. As a matter of fact, we use here the results of \cite{Melo:2015wpa} where spectra of pions, kaons, (anti)protons, and $\Lambda$'s were fitted simultaneously. For later simulations we use the centrality classes 0-5\%, 30-40\%, 50-60\%. The transverse size parameter $R_0$ is taken from  \cite{Chatterjee:2014lfa} and $s_0$ is determined so that the normalisation of the $p_t$ spectra comes out correct. The adopted parameters are listed in Table~\ref{table:Sampler_parameters}. 
%
\begin{table}[b]
\caption{The  parameters used for the generation of pions, protons, neutrons and direct deuterons from the Hadron Sampler for  centrality classes $0-5 \%$, $30-40 \%$ and $50-60 \%$.} 
\label{table:Sampler_parameters}
\begin{tabular}{c c c c c c c }      
\hline           
		 centrality& $T [\mathrm{MeV}]$ & $\eta_f$ & $R_0 [\mathrm{fm}]$ & $s_0 [\mathrm{fm/c}]$ & $a_2$ & $\rho_2$\\ 
		\hline\hline
		0-5\%  & $95$ & $0.98$ & $15.0$ & $21 \pm 2$ & $0.016$ & $0.008$ \\ 
        \hline
		30-40\%  & $106$ & $0.91$ & $10.0$ & $9 \pm 1$ & $0.085$ & $0.03$  \\ 
		\hline
		50-60\%  & $118$ & $0.80$ & $6.0$ & $6 \pm 0.5$ & $0.15$ & $0.02$  \\ 
		\hline
\end{tabular}
\end{table}
%

The differential $v_2$ then depends on spatial anisotropy parameter $a_2$ and flow anisotropy $\rho_2$, This dependence is also different for protons and for pions \cite{Tomasik:2004bn}. Hence, the combination of proton and pion $v_2(p_t)$ allows us to set both $a_2$ and $\rho_2$ uniquely. 

\begin{figure}[t]
	\includegraphics[width=0.45\textwidth, angle=0]{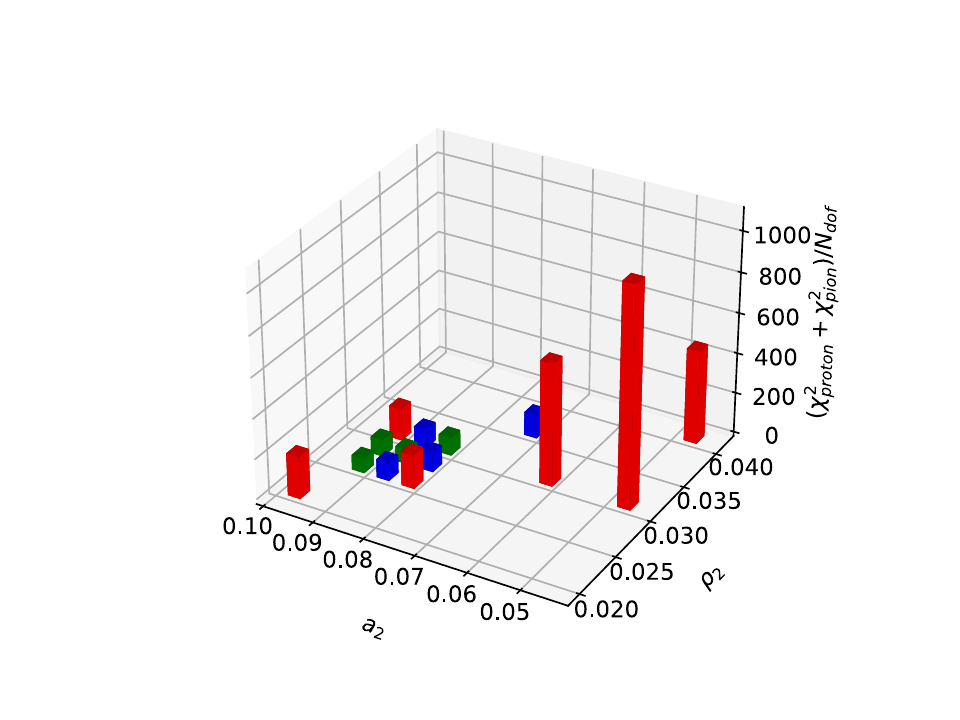}
	\caption{The $\chi^2$ from comparing theoretical results for various values of $a_2$ and $\rho_2$ with experimental data on $v_2(p_t)$ of protons and pions for the centrality class  30-40\%. \label{f:chimap} }
\end{figure}

Figure~\ref{f:chimap} illustrates for centrality 30-40\% the $\chi^2$ that we have obtained when we compared a few calculated $v_2(p_t)$ of protons and pions for various values of $a_2$ and $\rho_2$ with data. 
%
%
Their anticorrelation is observed but we see that a best fit can be obtained. 

At this point  we observed that we are unable to reproduce $v_2(p_t)$ reasonably when we stick to the assumption that the freeze-out hypersurface is set by constant $\tau$ and does not depend on $r$. The reason is that $v_2(p_t)$ requires rather strong elliptic flow. Since $\rho_0$ parametrizes the transverse {\em rapidity}, the parts of the fireball at the edges move with velocities close to that of the light. To see a sizeable effect on the elliptic flow we need a variation in the transverse expansion {\em velocity}. Close to the edge, the variation is rather limited since the expansion velocity is close to $c$ and must remain below $c$, The effect of expansion velocity variation is enhanced  if the freeze-out hypersurface bends downwards in the $\tau$-$r$ plane, i.e., regions which are farther from the longitudinal symmetry axis freeze-out earlier\footnote{The scenario is sometimes referred to as ``burning log''.}.
Technically, this is expressed by the value $s_2 = -0.02$~fm$^{-2}$. The obtained anisotropy parameters for centrality classes in which we calculated $v_2(p_t)$ are listed in Table~\ref{table:Sampler_parameters}.

\begin{figure}[t]
	\includegraphics[width=0.49\textwidth, angle=0]{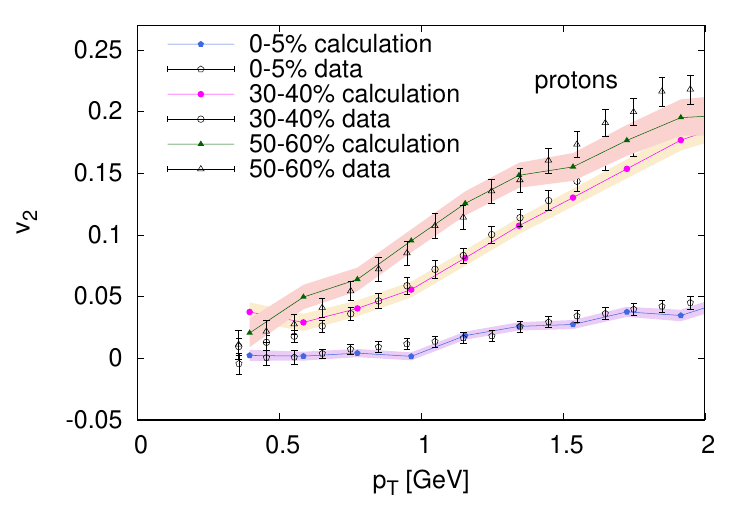}
	\caption{Elliptic flow of protons for centrality classes $0-5 \%$, $30-40 \%$, and $50-60 \%$ compared to data from Pb+Pb collisions at  $\sqrt{s_{NN}} = 2.76 ~\mathrm{TeV}$ by the ALICE collaboration. \label{f:v2prot} }
\end{figure}

\begin{figure}[t]
	\includegraphics[width=0.49\textwidth, angle=0]{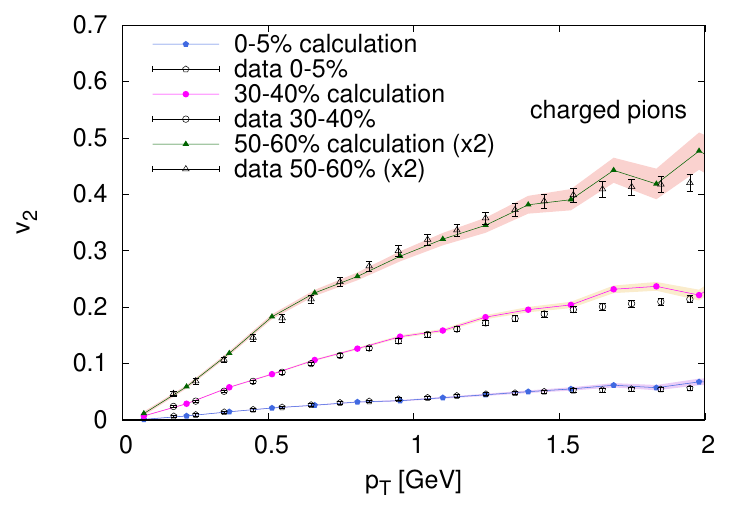}
	\caption{Elliptic flow of pions for centrality classes $0-5 \%$, $30-40 \%$, and $50-60 \%$  compared to data from Pb+Pb collisions at  $\sqrt{s_{NN}} = 2.76 ~\mathrm{TeV}$ by the ALICE collaboration. \label{f:v2pi} }
\end{figure}

The resulting $v_2(p_t)$ for protons and pions are plotted together with the experimental data in Figures
\ref{f:v2prot} and \ref{f:v2pi}, respectively. 
%
%
This demonstrates that we tuned our parametrisation well in order to reproduce the second-order anisotropy. In this paper we determined the $v_2$  with the event-plane method. 

\begin{figure}[t]
\includegraphics[width=0.49\textwidth, angle=0]{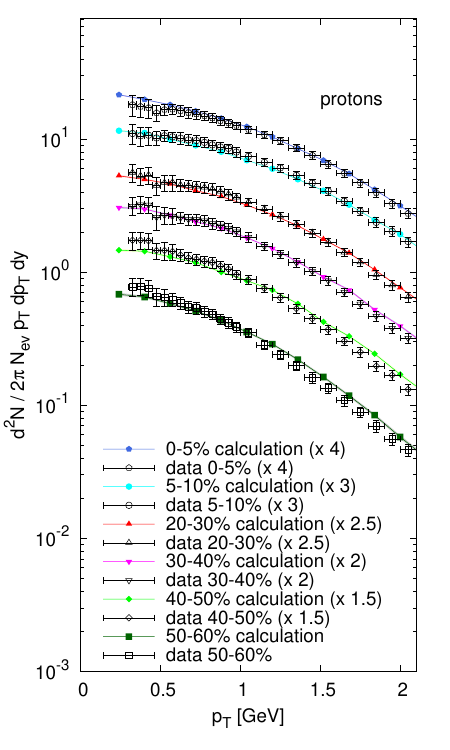}
\caption{Transverse momentum spectra of protons for centrality classes $0-5 \%$, $5-10 \%$, $20-30 \%$, $30-40 \%$, $40-50 \%$, and $50-60 \%$ compared to data from Pb+Pb collisions at  $\sqrt{s_{NN}} = 2.76 ~\mathrm{TeV}$ by the ALICE collaboration.
\label{f:spec_prot} }
\end{figure}

\begin{figure}[t]
\includegraphics[width=0.49\textwidth, angle=0]{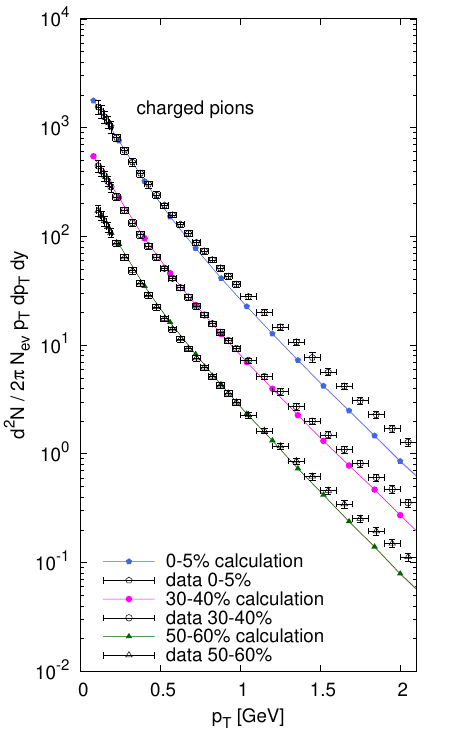}
	\caption{Transverse momentum spectra of pions for centrality classes $0-5 \%$, $30-40 \%$, and $50-60 \%$ compared to data from Pb+Pb collisions at  $\sqrt{s_{NN}} = 2.76 ~\mathrm{TeV}$ by the ALICE collaboration. 
\label{f:spec_pi} }
\end{figure}

The transverse momentum spectra of protons are shown in Fig.~\ref{f:spec_prot}
%
%
and those of pions in Fig.~\ref{f:spec_pi}. 
%
%
Since the deuteron $p_t$ spectra are measured in wider centrality classes, we will simulate those by merging the results from both smaller centrality classes (e.g., 20-30\% and 30-40\% will be merged into 20-40\% centrality class). To this end, we had to simulate also those classes which we do not use for the $v_2$ analysis. Figure~\ref{f:spec_prot} shows all resulting proton spectra. The parameters were determined in the same way as for the other centrality classes and are listed in Table~\ref{table:Sampler_parameters2}.
%
\begin{table}[b]
\caption{The  parameters used for the generation of pions, protons, neutrons and direct deuterons from the Hadron Sampler for  centrality classes $5-10 \%$, $20-30 \%$ and $40-50 \%$.} 
\label{table:Sampler_parameters2}
\begin{tabular}{c c c c c }     
\hline            
 centrality& $T [\mathrm{MeV}]$ & $\rho_0$ & $R_0 [\mathrm{fm}]$ & $s_0 [\mathrm{fm/c}]$ \\ 
 \hline\hline
5-10\%  & $97$ & $0.97$ & $13.0$ & $17$  \\ 
20-30 & $101$ & $0.94$ & $11.0$ & $11$  \\ 
40-50  & $112$ & $0.86$ & $7.0$ & $7$  \\ 
\hline
\end{tabular}
\end{table}


\section{Results and discussion}
\label{s:res}

After having constructed and calibrated the model it is now straightforward to simulate deuteron production. Thermal production is obtained just by using the mass and degeneracy factors of deuterons in the Cooper-Frye formula, Eq.~(\ref{e:CF}). Coalescence was described in subsection~\ref{ss:coal}. The maximum distance of proton and neutron to form a deuteron was chosen $\Delta r \le 3.5$~fm, in accord with the investigation \cite{Sombun:2018yqh}. The correct yield is then obtained with $\Delta p \le 0.26$~GeV. 

For each centrality class, $10^6$ events were initiated in the simulation chain. About 10\% of them did not finish due to computational issues. 

\begin{figure}[t]
\includegraphics[width=0.48\textwidth, angle=0]{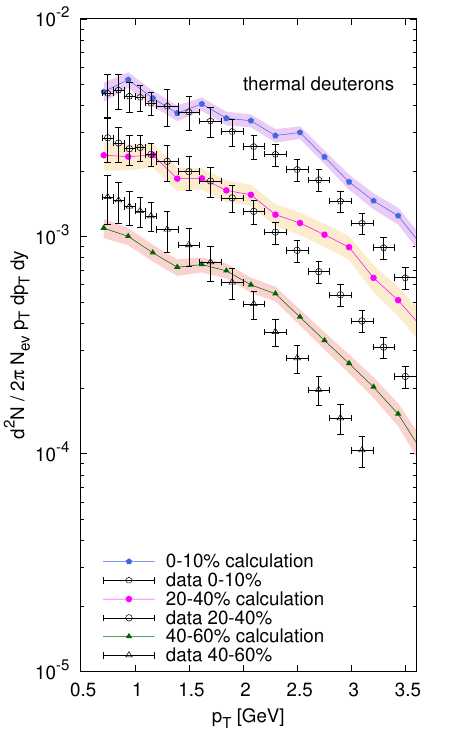}
\caption{Transverse momentum spectra of thermal deuterons for centrality classes $0-10 \%$, $20-40 \%$, and $40-60 \%$ compared with data from Pb+Pb collisions $\sqrt{s_{NN}} = 2.76 ~\mathrm{TeV}$ as measured by the ALICE collaboration. 
\label{f:d-spec-therm} }
\end{figure}

\begin{figure}[t]
	\centering	
	\includegraphics[width=0.48\textwidth, angle=0]{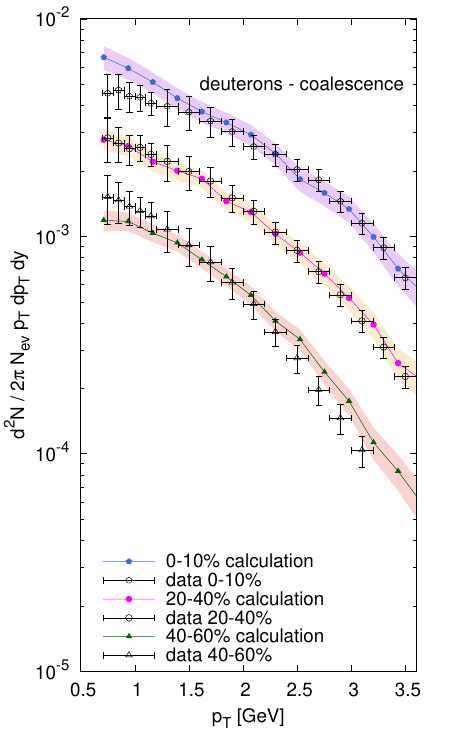}
	\caption{Transverse momentum spectra of deuterons from coalescence for centrality classes $0-10 \%$, $20-40 \%$, and $40-60 \%$ compared with data from Pb+Pb collisions $\sqrt{s_{NN}} = 2.76 ~\mathrm{TeV}$ as measured by the ALICE collaboration. 
\label{f:d-spec-coal} }
\end{figure}

In Figs.~\ref{f:d-spec-therm} and \ref{f:d-spec-coal} we present the transverse momentum spectra of deuterons produced thermally and by coalescence, respectively. 
%
%
At all investigated centralities, both models reproduce the yield (absolute normalisation of the spectra) and also the shape of the spectrum. Some discrepancies appear in thermally produced spectra, especially in non-central collisions, where the simulated spectra are flatter than data. The similarities of results can be understood, since the product of $f_p$ and $f_n$ in Eq.~(\ref{e:Dspect}) gives approximately the emission function of deuterons in the thermal model. The less detailed observable one looks at, the better is this rough argument. 

\begin{figure}[th]
\includegraphics[width=0.48\textwidth, angle=0]{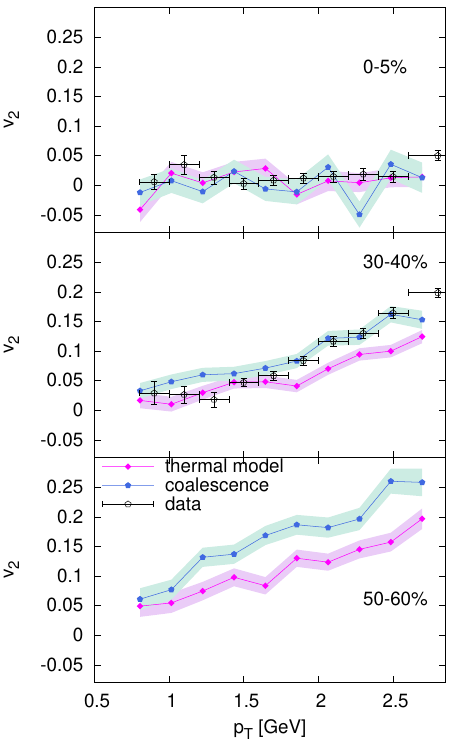}
\caption{Elliptic flow coefficient $v_2$ calculated for  thermal production as well as  coalescence for centrality classes $0-5 \%$, $30-40 \%$, and $50-60 \%$, and compared to data from Pb+P collisions at  $\sqrt{s_{NN}} = 2.76 ~\mathrm{TeV}$, as measured by the ALICE collaboration. 
\label{f:v2-deut} }
\end{figure}

Therefore, we look at the more detailed observable and examine the elliptic flow. In Fig.~\ref{f:v2-deut}
%
%
the $p_t$ dependence of $v_2$ is plotted. Experimental data are compared to both models used here. Recall that there in no room for tuning at this stage, since all parameters have been set by the comparison  to protons and pions and the last two parameters have been fixed by deuteron spectra. Now, we  see a clear difference between coalescence and thermal production. 
Generally, in non-central collisions coalescence leads to higher elliptic flow than the statistical production. In the 20-30\%\ centrality class it is the preferred model, particularly for $p_t$ above 1.5~GeV. In central collisions, elliptic flow is almost zero and the simulation results are strongly affected by statistical fluctuations. 

As we wanted to confirm our hypothesis that in more peripheral collisions coalescence may be more sensitive to the azimuthal variation of the homogeneity region, we looked at the 50-60\% centrality class. This is just one decile beyond the coverage by experimental data and we suppose that it could be covered with better experimental statistics. Here, the difference between coalescence and the thermal model prediction is clear, beyond the applied error bars. 

\begin{figure}[t]
\includegraphics[width=0.48\textwidth, angle=0]{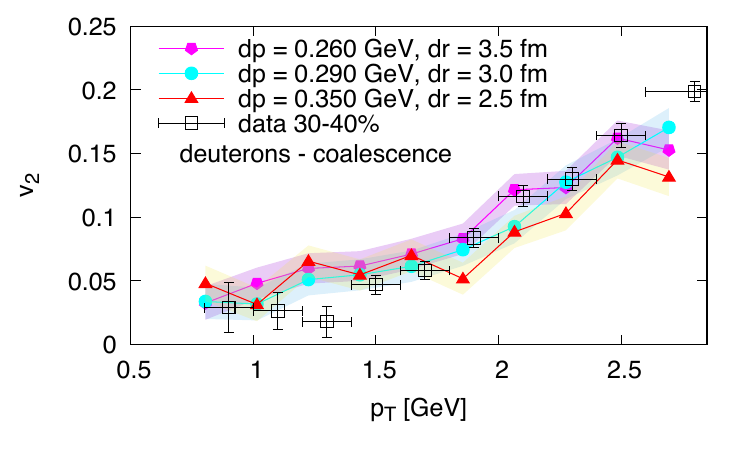}
\caption{Elliptic flow of deuterons from coalescence for different values of $\Delta r$ and $\Delta p$, compared to experimental data from Pb+P collisions at  $\sqrt{s_{NN}} = 2.76 ~\mathrm{TeV}$, as measured by the ALICE collaboration.  
\label{f:v2-deut-dr} }
\end{figure}

The elliptic flow---being actually the measure of transverse flow anisotropy---is larger from coalescence, when the size of the homogeneity region from which the deuterons come varies just around the size of the deuteron. The yield from coalescence is very sensitive to the interplay of these two sizes. Hence, one might expect that the anisotropy will be very sensitive to the spatial deuteron size, which is effectively set by $\Delta r$, here. Such a dependence might limit the robustness of our conclusions, since $\Delta r$ is one of our model parameters.  In order to test this robustness, we run simulations in which we lowered the value of $\Delta r$ and increased $\Delta p$ so that the normalisation of the spectra stays unchanged. We decreased $\Delta r$ by as much as 1~fm, but stayed conservative and did not go for lower values. The results for $v_2$ are plotted in Fig.~\ref{f:v2-deut-dr} for the 30-40\% centrality class. 
%
%
As the size of the deuterons is effectively lowered, they become less sensitive to the variation of the homogeneity lengths and the anisotropy of the deuteron distribution---measured by $v_2$---decreases. With the current variation of $\Delta r$ the decrease is such that all obtained curves for $v_2$ stay close to experimental data.  We stick to our conclusion that coalescence leads to larger $v_2$ than the thermal production. 

\begin{figure}[t]
\includegraphics[width=0.5\textwidth, angle=0]{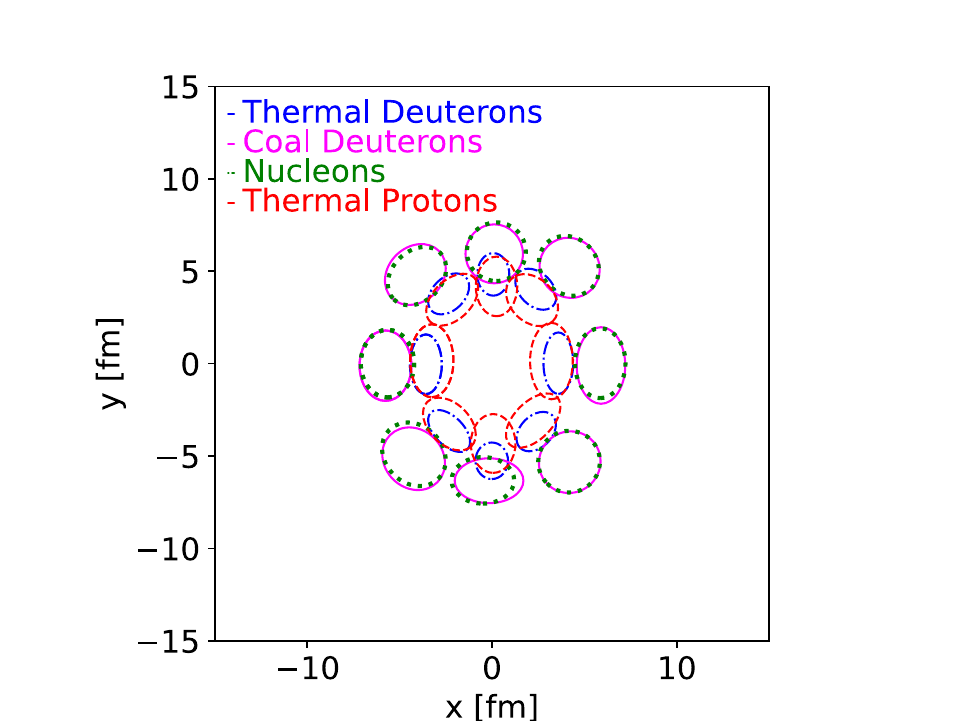}
\caption{Variances of the homogeneity regions that produce protons with $p_t \approx 1$~GeV and deuterons with $p_t\approx2$~GeV, divided into eight bins by azimuthal angle of the momentum. Centrality class is 50-60\%. The ellipses are centered around mean position of emission of given sort of particles and their half-axes are given by the variances of the emission points. We plot thermal protons (red dashed line), thermal deuterons (blue dash-dotted line), all nucleons that merge into deuterons via coalescence (green dotted line), and deuterons from coalescence (purple solid line). 
\label{f:ellipses} }
\end{figure}

We stressed several times, already, that for the yield from coalescence the size of the effective homogeneity region (a.k.a.\ homogeneity length) in comparison to the deuteron size is decisive. In order to explore this issue, we want to have a look at the homogeneity lengths in Fig.~\ref{f:ellipses}. 
%
%
We examine the centrality class 50-60\%, so that the anisotropy is clearly visible and focus on deuterons with $p_t$ from the interval $2.0\pm 0.2$~GeV and protons with $p_t$ from the interval $1.0\pm0.1$~GeV. They are divided into 8 bins of azimuthal angle of their momentum. For each type of particles we recorded the positions from which they were emitted. For deuterons from coalescence this means that we record the position where we calculate that the nucleon pair merged into the deuteron. From the obtained samples of emission points we determine the mean positions and the variances in the outward direction and in the direction perpendicular to it. Then, we draw ellipses with centres given by the mean emission position and the half-axes given by the variances. Thus, the ellipses represent the information about the homogeneity regions and lengths. 

Figure~\ref{f:ellipses} reveals an important insight: the intuitive picture based on the thermal production from the fireball must be modified, because a part of the nucleons comes from decays of resonances. This shows up as a shift of the production region in the outward direction. For nucleons, the shift is about 2.5~fm. This shift has an impact on the emission region of the deuterons. Those from coalescence are produced also from nucleons that stem from resonance decays. Their emission region coincides with that of all nucleons. On the other hand, statistical production of deuterons does not include resonance decays. As a consequence, the source of statistically produced deuterons must fit into the thermal fireball. Due to higher mass of the deuterons with respect to protons, the thermal source of deuterons is smaller than that of protons and it is pushed further towards the edge of the fireball. It might be interesting to explore the sizes of the emitting regions with the help of femtoscopy, as recently proposed in \cite{Mrowczynski:2019yrr,Mrowczynski:2021bzy}. Figure~\ref{f:ellipses} confirms that such a study may indeed also help with the discrimination of potential deuteron production mechanism.


\section{Conclusions}
\label{s:conc}

We conclude that the second-order azimuthal anisotropy of deuterons, a.k.a.\ elliptic flow $v_2$, is larger in case of coalescence than it is for the direct statistical production of deuterons. Our claim is based upon simulation with carefully calibrated model that is fixed by reproducing the available data on proton and pion transverse momentum spectra and $v_2$.

Let us recall the important features of our approach which support the robustness of our conclusions: 
\begin{itemize}
\item 
Our simulations include the decays of resonances. 
\item
We employ a fully developed parametrization of the emission and overcome numerical difficulties with integrating of the emission function by Monte Carlo sampling of the events, so that all details of source shape and flow are included. 
\item 
We acknowledge that there are two potential causes for azimuthal anisotropy of hadron distribution: anisotropic shape of the fireball and the anisotropy of its expansion. We include them both and tune them via comparison to proton and pion $v_2(p_t)$. 
\item 
We insist on reproduction of proton (nucleon) and pion data for $p_t$-spectra and $v_2$. In the hadronic fireball, these are the most abundant species and the former interact through the exchange of the latter. Hence, they form together the bulk of the fireball and deuterons can be produced when this bulk of nucleons and pions decouples.
\end{itemize}

On the other hand, we remained on a simpler level with modelling of the deuteron wave function. In general, any model wave function of the deuteron will include its size and so we expect that our results will stay qualitatively valid with any of them. The study \cite{Mahlein:2023fmx} has shown the sensitivity of the calculated deuteron yield to the model for the wave function that is used, albeit with rather simple modelling of the source. It will be very interesting, though beyond our scope in this paper, to try other models for the deuteron wave function with this approach.

Note that other use of the blast-wave model to reproduce elliptic flow has  been documented in literature \cite{Oh:2009gx,Yin:2017qhg,Zhu:2017zlb}. To achieve this,  azimuthal variation of the transverse expansion velocity of the bulk was implemented as depending on the $p_t$ of the produced particle. This may mimic continuous freeze-out as it naturally occurs in transport models. In our work, we stayed strictly within the framework of the blast-wave model which assumes common freeze-out for all hadrons. We showed that it is possible to reproduce deuteron spectra and elliptic flow within these constraints.

The method of using $v_2$ for the distinction between coalescence and statistical production makes use of the homogeneity lengths that vary with the azimuthal angle of the emission. It may work if the homogeneity lengths are comparable or smaller than the size of the deuteron, otherwise we would expect similar $v_2$ from both production mechanisms. Hence, the method should work best in collisions with larger impact parameter when the spatial anisotropy can be largest. Whether it works as well in collisions of smaller systems, when the homogeneity lengths are always  smaller than the typical size of the deuteron wave function, remains to be studied. 

It may appear as an attractive idea to apply the same kind of study on $v_2$ of larger clusters, since their wave functions have different sizes and may be more sensitive to the azimuthal variation of the homogeneity lengths. 
On the other hand, the applicability of the method will be limited by the available statistics. 
This idea remains to be studied, as well.


\section*{Acknowledgments}

This project is supported by the Czech Science Foundation (GA\v{C}R) under No 22-25026S.
BT acknowledges support from VEGA 1/0521/22.
Computational resources were provided by the e-INFRA CZ project (ID:90254),
supported by the Ministry of Education, Youth and Sports of the Czech Republic.
We thank Niklas G\"otz for the assistance with setting SMASH. 


\appendix

\section{The modified distribution of particle momenta}
\label{a:PTM}

Here we explain the symbols that appear in the modified momentum distribution displayed in Eq.~(\ref{eq:5.9}). Note that it accounts for  the departure of the distribution from the perfect-fluid equilibrium case due non-vanishing  gradients. We chose this version of correction because it is more robust than the method of adding correction terms \cite{Teaney:2003kp} which may make the resulting distribution negative for too large momenta. 
We closely follow here the formalism of \cite{McNelis:2021acu}. 

For later reference, we  start with some definitions and preliminaries. 


\subsection{Definitions}
\label{a:defs}

To keep the notation less cumbersome, we  use $a\cdot b$ for the scalar product of two four-vectors ($a\cdot b = a_\mu b^\mu$) and also for multiplication with tensors 
(e.g., $a\cdot t \cdot b = t^{\mu\nu}a_\mu b_\nu$).
Transverse projector to velocity is defined as
\begin{equation}
    \Delta^{\mu\nu} = g^{\mu\nu} - u^\mu u^\nu\,  , 
\end{equation}
and the  projector for tensors as
\begin{equation}
    \Delta^{\mu\nu}_{\alpha\beta} =\frac{1}{2} \left ( \Delta^\mu_\alpha \Delta^\nu_\beta + \Delta^\mu_\beta \Delta^\nu_\alpha \right ) - \frac{1}{3} \Delta^{\mu\nu}\Delta_{\alpha\beta}\, .
\end{equation}

With its help, one can define the velocity shear tensor
\begin{equation}
    \sigma_{\mu\nu} = \Delta^{\alpha\beta}_{\mu\nu} \partial_\beta u_\alpha\,  .
\end{equation}

The fluid Local Rest Frame (LRF) is the one in which spatial components of $u^\mu$ vanish. In LRF, the three spatial dimensions can be spanned by three unit vectors $X_i$. These vectors can be boosted to any other reference frame and their components will be $X^\mu_i$. With their help, one can project out from four-vectors their components, which make up the spatial components in LRF
\begin{equation}
    w_i = X^\mu_i w_\mu\, ,
\end{equation}
and similarly for tensors
\begin{equation}
    t_{ij} = X^\mu_i X^\nu_j t_{\mu\nu}\,  .
\end{equation}

The locally equilibrated distribution function for species $n$ is the standard Bose-Einstein of Fermi-Dirac distribution 
\begin{equation}
    \feqn (x,p) = \frac{g_n}{\exp\left (  \frac{p\cdot u(x)}{T(x)}  - \alpha_n(x) \right ) + \Theta_n}\, .
\end{equation}
Recall that $g_n$ is the spin degeneracy, $\alpha_n(x) = \mu_n(x)/T(x)$, and $\Theta_n = 1\, (-1)$ for fermions (bosons). One also defines
\begin{equation}
    \feqnb = 1- g_n^{-1} \Theta_n \feqn\,  .
\end{equation}
A shorthand is introduced for thermal integrals of species $n$
\begin{equation}
    J_{kq,n} = \int\frac{d^3p}{(2\pi)^3 E} \frac{(u\cdot p)^{k-2q} (-p\cdot \Delta \cdot p)^q}{(2q+1)!!} \feqn\feqnb \, ,
\end{equation}
and they are summed over all species in 
\begin{eqnarray}
    \cJ_{kq} & = & \sum_n J_{kq,n} \,  ,\\
    \cN_{kq} & = & \sum_n b_n J_{kq,n} \,   , \\
    \cM_{kq} & = & \sum_n b^2_n J_{kq,n} \,  .
\end{eqnarray}

The energy density and the net baryon density in equilibrium are 
\begin{eqnarray}
    {\cal E} & = & \sum_n \int\frac{d^3p}{(2\pi)^3 E} (u\cdot p)^2 \feqn\, , \\
    n_B & = & \sum_n b_n \int\frac{d^3p}{(2\pi)^3 E} (u\cdot p) \feqn\, ,
\end{eqnarray}
and the equilibrium pressure
\begin{equation}
    \Peq = \frac{1}{3} \sum_n \int \frac{d^3p}{(2\pi)^3 E} (-p\cdot \Delta \cdot p) \feqn\, .
\end{equation}

We shall also use the following temperature-dependent coefficients
\begin{eqnarray}
    \cG & = & T \frac{({\cal E} + \Peq) \cN_{20} - n_B \cJ_{30}}{\cJ_{30} \cM_{10} - \cN_{20}^2}\,  ,\\
    \cF & = & T^2 \frac{n_B \cN_{20}- ({\cal E} + \Peq)\cM_{10}}{\cJ_{30}\cM_{10} - \cN{20}^2}\,  .
\end{eqnarray}

\subsection{The formula for the momentum distribution}
\label{a:formul}

Now we can explain the formula Eq.~(\ref{eq:5.9}), which we repeat here for convenience: 
\begin{equation}
\label{ae:PTM}
    f_{eq,n}^{PTM} = \dfrac{Z_n g_n}{\exp \left[ \dfrac{\sqrt{\boldsymbol{p}^{\prime 2} + m_n^{2}}}{T + \beta_{\Pi}^{-1} \Pi \cF} - b_n \left( \alpha_B + \dfrac{\Pi \cG}{\beta_{\Pi}}\right) \right] + \Theta_n},
\end{equation}
The thermal distribution works with the momentum $\boldsymbol{p}'$. It is related to the momentum in  Local Rest Frame (LRF) $\boldsymbol{p}$ through 
\begin{equation}
    p_i = A_{ij}p_j^\prime - q_i \sqrt{{\boldsymbol{p}'}^2 + m_n^2  } + b_n T a_i\, 
\end{equation}
where
\begin{eqnarray}
A_{ij} & = & \left ( 1 + \frac{\Pi}{3\beta_\Pi}   \right )\delta_{ij} + \frac{\pi_{ij}}{2\beta_\pi}\,  ,\\
q_i & = & \frac{V_{B,i} n_B T}{\beta_V({\cal{E}} + {\cal{P}}_{eq})}\, ,\\
a_i & = & \frac{V_{B,i}}{\beta_V}    \, .
\end{eqnarray}
For the bulk viscous pressure and the shear stress tensor we assume the Navier-Stokes form
\begin{eqnarray}
    \Pi & = & -\zeta \partial_\mu u^\mu \, ,\\
    \pi^{\mu\nu} & = & 2\eta\sigma^{\mu\nu} \, , 
\end{eqnarray}
nevertheless, in our application we set the bulk viscosity identically 0. For the shear viscosity we assume $\eta/s = 0.2$.
The LRF spatial components of shear stress are $\pi_{ij} = X_i^\mu X_j^\nu \pi_{\mu\nu}$.
The Navier-Stokes expression for the baryon current would be 
\begin{equation}
    V_B^\mu = \kappa_B \Delta^{\mu\nu}\partial_\nu \left ( \frac{\mu(x)}{T(x)} \right )\,  ,
\end{equation}
with $\kappa_B = T\sigma_B$ and $\sigma_B$ is the baryon conductivity. Then, $V_{B,i} = X^\mu_i V_{B,\mu}$ is its spatial component in the LRF. Nevertheless, our density profile and temperature do not vary with the coordinates and so the baryon current vanishes. 

Out of the ratios of bulk viscosity, baryon diffusion coefficient, and shear viscosity to the relaxation time---$\beta_\Pi$, $\beta_V$, and $\beta_\pi$---we thus do not need the former two. The third one is determined as 
\begin{equation}
    \beta_\pi = \frac{\cJ_{32}}{T}\,  .
\end{equation}

The factor  $Z_n$ normalizes the distribution to fix the particle density. It is chosen as \cite{McNelis:2021acu}
\begin{equation}
    Z_n = \frac{1}{\mathrm{det}A} \frac{n_n^{(1)}}{n_{\mathrm{eq},n}(T+\beta_\pi^{-1}\Pi\cF,\alpha_B + \beta^{-1}_\Pi\Pi \cG)}\, ,
\end{equation}
where $n_{\mathrm{eq},n}(T,\alpha)$ is the equilibrium particle density and 
\begin{equation}
    n_n^{(1)} = n_{\mathrm{eq},n} + \frac{\Pi}{\beta_\Pi} \left ( n_{\mathrm{eq},n} + b_n J_{10,n}\cG + \frac{J_{20,n}\cF}{T^2}  \right )\,  .
\end{equation}
Nevertheless, since we do not have the bulk viscous pressure in our simulations, the normalisation simplifies to just 
\begin{equation}
    Z_n = \frac{1}{\mathrm{det}A} \, .
\end{equation}

\end{document}